\documentclass[floatfix,%
 reprint,%
 amssymb, amsmath,%
 aip,jcp,
frontmatterverbose,
]{revtex4-2}

\usepackage{amsmath,amssymb,amsfonts,mathtools}
\usepackage{multirow}
\usepackage{float}
\usepackage{csquotes}
\usepackage{dsfont}
\usepackage[mathscr]{euscript}
\usepackage{natbib}

\usepackage{enumitem}
\setlist{nosep} 

\usepackage{graphicx}
\usepackage{bm}%
\usepackage{longtable}
\usepackage{supertabular}
\usepackage{rotating}
\usepackage[colorlinks=false,linkcolor=blue]{hyperref}
\usepackage[normalem]{ulem}
\usepackage{ifthen}
\newboolean{includeMQDT}
\setboolean{includeMQDT}{false}  

\usepackage[dvipsnames]{xcolor}

\usepackage{dcolumn,ulem}  
\usepackage{xspace}

\begin{document}
\newcommand{\pcut}[1]{{\color{blue}{\sout{#1}}}}   
\newcommand{\Vel}{\ensuremath{V_\mathrm{el}}}
\newcommand{\SH}{\ensuremath{\mathrm{SH}}}
\newcommand{\SHplus}{\ensuremath{\mathrm{SH}^+}}
\newcommand{\il}{{\it et~al.}}
\newcommand{\Hdia}{\ensuremath{\mathbf{H}_\mathrm{dia}}}

\author{Braden M. Weight}
\affiliation{Department of Physics and Astronomy, University of Rochester, Rochester, NY 14627, U.S.A.}
\author{Arkajit Mandal}
\affiliation{Department of Chemistry, University of Rochester, Rochester, NY 14627, U.S.A.}
\author{Pengfei Huo}
\email{pengfei.huo@rochester.edu}
\affiliation{Department of Chemistry, University of Rochester, Rochester, NY 14627, U.S.A.}%
\affiliation{The Institute of Optics, Hajim School of Engineering, University of Rochester, Rochester, New York, 14627}
\title{\large \textit{Ab-initio} Symmetric Quasi-Classical Approach to Investigate Molecular Tully Models}


\begin{abstract}
We perform on-the-fly non-adiabatic molecular dynamics simulations using the symmetrical quasi-classical (SQC) approach with the recently suggested molecular Tully models: ethylene and fulvene. We attempt to provide benchmarks of the SQC methods using both the square and the triangle windowing schemes as well as the recently proposed electronic zero-point-energy correction scheme (so-called the $\gamma$ correction). We use the quasi-diabatic propagation scheme to directly interface the diabatic SQC methods with adiabatic electronic structure calculations. Our results showcase the drastic improvement of the accuracy by using the trajectory-adjusted $\gamma$-corrections, which outperform the widely used trajectory surface hopping method with decoherence corrections. These calculations provide useful and non-trivial tests to systematically investigate the numerical performance of various diabatic quantum dynamics approaches, going beyond simple diabatic model systems that have been used as the major workhorse in the quantum dynamics field. At the same time, these available benchmark studies will also likely foster the development of new quantum dynamics approaches based on these techniques.
\end{abstract}

\maketitle
\section{Introduction}

Simulating on-the-fly non-adiabatic quantum dynamics in molecular systems remains a central challenge in modern theoretical chemistry, despite the impressive progress made in the past several decades\cite{TullyJCP1990,Ben-NunJPCA2000,MichaJPCA1999,BonnellaJCP01,WorthFD2004,AnanthJCP2007,HuoJCP2011,PlasserJCP2012,SaitaJCP2012,DmitryJCP2014, CottonJCP2013,CottonJPCA2013,AnanthJCP2013, HsiehJCP2013,RichardsonJCP2013,MenzeleevJCP2014, MeekJCPL2014, SubotnikARPC2016,NelsonCP2016,JainJCTC2016,WaltersJCP2016,Pal2016JCTC, ChowdhuryJCP2017,Pal2016JCTC,CurchodChemRev2018,CrespoChemRev2018,MulvihillJCP2019,CottonJCP2019_2,Runeson2020}. The two main components for performing an on-the-fly quantum dynamics simulation are (i) obtaining accurate electronic structure information and (ii) using it to propagate the coupled motion of nuclear and electronic degrees of freedom (DOF) in an efficient manner\cite{TullyJCP2012}. Mixed quantum classical (MQC) approaches such as the fewest-switches surface hopping\cite{TullyJCP1990} (FSSH) and the mean-field Ehrenfest\cite{KabPRE2002} (MFE) approach, which uses the outputs of electronic structure methods to evolve the electronic subsystem quantum mechanically and nuclear DOFs classically, has remained popular for simulating on-the-fly quantum dynamics. However, the inherent mixed-quantum classical approximation in these approaches can lead to the break-down of detail balance\cite{ParandekarJCTC2006}, the artificial creation of  electronic coherence\cite{SubotnikARPC2016} or incorrect
chemical kinetics\cite{SubotnikARPC2016}. 

In response to these deficiencies, a wide range of non-adiabatic dynamics approaches have been developed in the diabatic representation, some of which include partial linearized density matrix\cite{HuoJCP2011,HuoMP2012} (PLDM), symmetrical quasi-classical\cite{CottonJCP2013,CottonJPCA2013} (SQC), state-dependent ring polymer molecular dynamics\cite{RichardsonJCP2013, AnanthJCP2013,ChowdhuryJCP2017}, quantum-classical path integral (QCPI) approach\cite{LambertJCP2012_1,LambertJCP2012_2,BanerjeeJPCB2013,MakriIJQM2015}, and the quantum classical Liouville equation (QCLE) dynamics.\cite{HsiehJCP2012,HsiehJCP2013}  In particular, the recently developed $\gamma$-SQC has been shown\cite{CottonJCP2019_2} to provide impressively accurate non-adiabatic photo-dissociation quantum dynamics with coupled Morse potentials through the adjusted zero point energy parameter of the mapping variables, thus appearing to be a promising method to simulate on-the-fly quantum dynamics of complex molecular systems. Testing these approaches with simple model systems becomes the major workhorse in the quantum dynamics field. What has been largely missing, on the other hand, are the calculations that go beyond simple diabatic models. However, reformulating these approaches from diabatic to adiabatic representation often requires non-trivial theoretical tasks and introduces new numerical complications.

In our recent works, we have developed the quasi-diabatic (QD) propagation scheme\cite{MandalJCTC2018,MandalJCP2018,MandalJPCA2019,SandovalJCP2018,ZhouJCPL2019} as a general strategy to seamlessly combine a diabatic quantum dynamics approach with the adiabatic outputs of an electronic structure method. The QD propagation scheme uses the adiabatic states with a reference nuclear geometry (the so-called ``crude adiabatic" states) as the local diabatic states during a short-time propagation and continuously updates the QD basis at each consecutive nuclear propagation step. In this propagation scheme, one does not construct a global diabatic representation but uses a sequence of local diabatic representations for each short-time segment to propagate quantum dynamics. Note that the quasi-diabatic propagation scheme \cite{MandalJCTC2018,MandalJCP2018,MandalJPCA2019,SandovalJCP2018,ZhouJCPL2019} should not be confused with the approximate diabatic representation which are also referred often as the QD representation in the literature.\cite{GuanJCP2019,ZhuJCP2012,YuchenJPCA2019}

In this work we use the QD propagation scheme to combine the $\gamma$-SQC approach with adiabatic outputs of the complete active space self consistent field (CAS-SCF) approach to perform on-the-fly non-adiabatic quantum dynamics in molecular systems. We directly simulate photo-excited non-adiabatic dynamics in two molecular systems, ethylene and fulvene. These molecular systems have been recently proposed\cite{IbelePCCP2020} to be analogues of the Tully curve-crossing models,\cite{TullyJCP1990} I and III, respectively. The original Tully models\cite{TullyJCP1990} explore nuances in excited state dynamics that are ubiquitous in `real' molecular systems to varying degrees of complexity, while only involving one degree of freedom, and have been extensively used to benchmark quantum dynamics approaches.\cite{AnanthJCP2007,DukeJCPL2015,HuoJCP2011,CottonJCP2013,ShakibJPCL2017,GaoJCTC2020} The molecular analogues of the Tully models, on the other hand, capture the basic physics of the original Tully models while concurrently showcasing complex dynamical features due to the coupled motion between the electronic and multiple nuclear DOF. These molecular systems serve as robust benchmarks offering complex non-adiabatic dynamics beyond one-dimensional, overly simplified model systems and are representative of typical molecular systems. 

Our numerical results demonstrate that the zero-point energy (ZPE) corrected SQC ($\gamma$-SQC ) improves the population dynamics compared to the original SQC approach, benchmarked against to the accurate but expensive {\it ab-initio} multiple spawning (AIMS) approach. Our numerical results also show that $\gamma$-SQC can outperform the state-of-the-art decoherence-corrected surface hopping (dTSH) approach. Overall, our results demonstrate the accuracy and applicability of the $\gamma$-SQC approach for {\it ab-initio} on-the-fly simulation enabled by the QD propagation scheme, opening up future opportunities for  simulating on-the-fly quantum dynamics of complex molecular systems.

\section{Theory}
The SQC approach\cite{CottonJCP2013,CottonJPCA2013} uses symmetrical window functions to sample electronic DOF at initial time and provides an estimate of the reduced density matrix at later times. It relies on the Meyer-Miller-Stock-Thoss (MMST) mapping Hamiltonian approach, which transforms the electronic degrees of freedom onto an effective set of singly-excited and fictitious classical harmonic oscillators.  It has been shown to provide accurate non-adiabatic dynamics in a wide range of model systems\cite{CottonJCP2013,CottonJCP2014,CottonJCTC2016,MillerFD2016}. Recently, the original SQC method was also combined with the QD propagation scheme\cite{SandovalJCP2018} for direct quantum dynamics simulation seamlessly using adiabatic electronic structure outputs.\cite{ZhouJCPL2019} Here, we briefly discuss the essential idea of the mapping Hamiltonian, the $\gamma$-SQC approach and the QD propagation scheme.

\subsection{Mapping Hamiltonian Formalism}
The Meyer-Miller-Stock-Thoss (MMST) mapping representation \cite{Meyera1979,StockPRL1997,ThossPRA1999} transforms the discrete electronic DOFs onto an effective set of fictitious, singly-excited classical harmonic oscillators, thus mapping the electronic non-adiabatic dynamics onto these oscillators' phase space motion.

The total molecular Hamiltonian in the diabatic representation is expressed as follows
\begin{equation}\label{Diabatic-Hamiltonian}
\hat H = \hat T + \sum_{ij} V_{ij}(\hat{\bf R})|i \rangle\langle j|, 
 \end{equation}
where $V_{ij}(\hat{\bf R})=\langle i|\hat{V}(\hat{\bf r},\hat {\bf R})|j\rangle$ are the matrix elements of the electronic Hamiltonian in the diabatic basis $\{|i\rangle\}$. Using the Meyer-Miller-Stock-Thoss\cite{MeyerJCP1979,StockPRL1997,ThossPRA1999} mapping representation, the discrete electronic states are transformed into continuous phase-space variables
\begin{equation} \label{eq:mmst}
|i\rangle\langle j| \rightarrow  {\hat a}_{i}^\dagger {\hat a}_{j},
\end{equation} 
where ${\hat a}^\dagger_{i} =({\hat q}_{i} - i{\hat p}_{i})/\sqrt{2}$ and ${\hat a}_{j} =({\hat q}_{j} + i{\hat p}_{j})/\sqrt{2}$. With this transformation, the molecular Hamiltonian in Eq.~\ref{Diabatic-Hamiltonian} is transformed into the following MMST mapping Hamiltonian
\begin{equation} \label{eq:map} 
\hat{H}_\mathrm{m}=\hat{T}+{\frac{1}{2}}\sum_{ij}V_{ij}(\hat{R})\left(\hat{p}_{i}\hat{p}_{j}+\hat{q}_{i}\hat{q}_{j}-2\gamma\delta_{ij}\right), 
\end{equation}
where $\gamma=0.5$ is the ZPE for the mapping harmonic oscillators. Historically, it is recognized as the Langer correction by Meyer and Miller\cite{MeyerJCP1979} for the quasi-classical description. Note, until Eq.~\ref{eq:map}, no approximations have been made. 

In the SQC approach, instead of evolving all DOF quantum mechanically, the coupled electronic-nuclear dynamics are propagated using the following classical Hamiltonian\cite{MillerFD2016}
\begin{equation} \label{eq:mapham} 
    {H}_\mathrm{m}={{{\bf P}^2}\over {2M}}+{\frac{1}{2}}\sum_{ij}V_{ij}(R)\left(p_{i}p_{j}+q_{i}q_{j}-2\gamma\delta_{ij}\right),
\end{equation}
where $\gamma$ is viewed as a parameter\cite{MillerFD2016} which specifies the ZPE of the mapping oscillators.\cite{UweJCP1999} 

Classical trajectories are generated based on Hamilton's equations of motion
\begin{eqnarray} \label{eq:mapeqn} 
    \dot q_{j} &=& \partial H_\mathrm{m}/ \partial p_{j};~~\dot p_{i} = -\partial H_\mathrm{m} / \partial q_{i}\\
    \dot {\bf R} &=& \partial H_\mathrm{m}/ \partial {\bf P};~~ \dot {\bf P}=-\partial H_\mathrm{m} / \partial {\bf R}= {\bf F}, 
\end{eqnarray}
with the nuclear force expressed as
\begin{equation}
\label{eq:force}
    {\bf F}=-{1\over 2}\sum_{ij}\nabla V_{ij}({\bf R})\big(p_{i}p_{j}+q_{i}q_{j}-2\gamma\delta_{ij}\big).
\end{equation}
Overall, the MMST mapping Hamiltonian provides a consistent classical footing for both electronic and nuclear DOFs, and the non-adiabatic transitions between electronic states are captured through the classical motion of the fictitious harmonic oscillators.

\subsection{Symmetric Quasi-Classical (SQC) Approach} \label{SEC:SQC}
To sample the electronic initial condition and estimate the population, the SQC approach uses the action-angle variables, $\{e_{j}, \theta_{j}\}$, which are related to the canonical mapping variables $\{p_{j}, q_{j}\}$ through 
\begin{equation}
    e_j = \frac{1}{2}\left(p_j^2 + q_j^2 \right);~~~\theta_j =-\tan^{-1}\left( \frac{p_i}{q_i}\right),
\end{equation}
and the inverse relations are
\begin{equation}
    q_j = \sqrt{2 e_j }\cos(\theta_j);~~p_j =-\sqrt{2 e_j}\sin(\theta_j),
\end{equation}
where $e_j$ is a positive-definite action variable introduced by Cotton and Miller that is directly proportional to the mapping variables' radius in action-space,\cite{CottonJCP2019_2} which allows for conceptual simplification (compared to the $n_j = e_j - \gamma$ action variable used in previous works\cite{SandovalJCP2018,CottonJCP2013,CottonJCP2016}) as it is independent of the ZPE parameter $\gamma$, which will be allowed to be state-dependent in subsequent sections of this work.

The SQC approach allows for the population of electronic state $|j\rangle$ to be evaluated as\cite{MillerFD2016}
\begin{align}\label{eqn:wignersqc}
    \rho_{jj}(t)&=\mathrm{Tr}_{\bf R}\left[\hat{\rho}_{R}|i\rangle\langle i|e^{i\hat{H}t/\hbar}|j\rangle\langle j|e^{-i\hat{H}t/\hbar}\right]\\
&\approx\frac{1}{\left(2\pi \hbar\right)^{N+M}}\int d\boldsymbol{\tau} \rho_\mathrm{W}({\bf P},{\bf R})W_i(\mathbf{e}(0))W_j(\mathbf{e}(t)),\nonumber
\end{align}
where $\hat{\rho}(0)=|i\rangle \langle i|\otimes \hat{\rho}_{R}$ is the initial density operator, $\rho_\mathrm{W}({\bf P},{\bf R})$ is the Wigner density of $\hat{\rho}_{R}$ operator that contains $\mathcal{N}$ nuclear DOFs, $\mathbf{e} = \{e_1,e_2,...,e_\mathcal{F}\}$ is the positive-definite action variable vector for $\mathcal{F}$ electronic states, $W_i(\mathbf{e})= \delta (e_i - (1+\gamma))\prod_{i \neq j} \delta(e_j-\gamma)$ are the Wigner transformed action variables,\cite{MillerJCP2016} and $d\boldsymbol{\tau}\equiv d{\bf P}\cdot d{\bf R}\cdot d\mathbf{e} \cdot d\boldsymbol{\theta}$. 

\begin{figure}
 \centering
  \begin{minipage}[h]{1.0\linewidth}
     \centering
 \includegraphics[width=\linewidth]{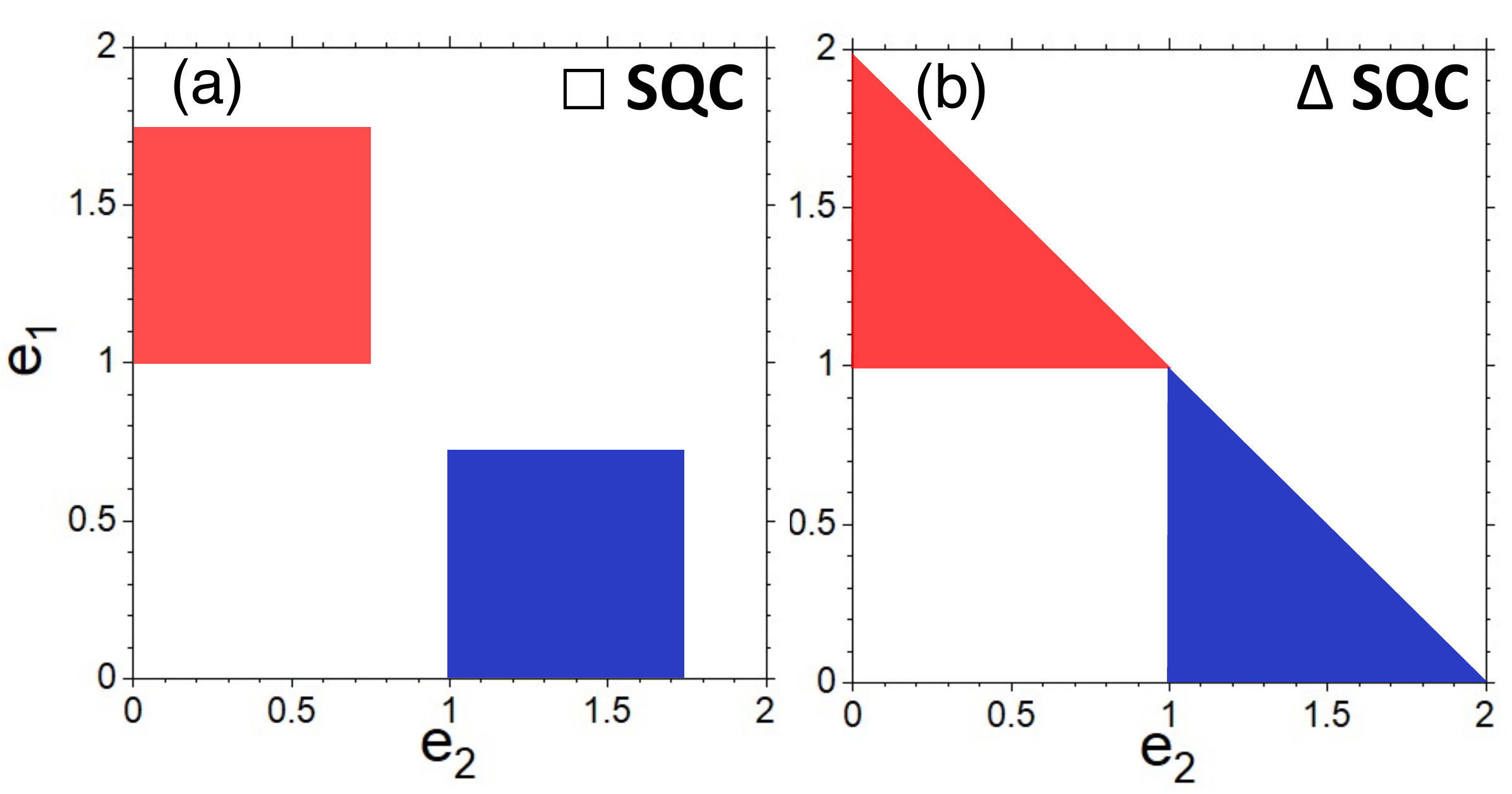}
       \end{minipage}%
   \caption{(a) Symmetric quasi-classical (a) square $\square$ and (b) triangle $\Delta$ window distributions depicted for a two-state projection involving state 1 (blue) and 2 (red). The action-space window distributions are depicted using the positive-definite action variables $\{e_k\}$, which are shifted quantities according to the corresponding zero-point energy parameter $\gamma$.}
\label{FIG1}
\end{figure}

For practical reasons, the above delta functions are artificially broadened using two well-explored distribution functions (i.e., square and triangle) that can be used to bin the resulting electronic action variables in action-space, depicted for any two-state projection in Fig. \ref{FIG1}a,b.\cite{MillerFD2016} The square distribution for an $\mathcal{F}$-state system is defined as\cite{CottonJCP2019_2}
\begin{equation}\label{EQ:SquareWindow}
    W_j({\bf e}) = w_{1}(e_j) \prod_{j'\ne j}^\mathcal{F} w_{0}(e_{j'}),
\end{equation}
where the function $w_{N}(e)$ is expressed as
\begin{equation}
    w_{N}(e) = 
    \begin{cases} 
      1, & 0 < e - N < 2 \ \times \ 0.366 \\
      0, & \mathrm{else}
   \end{cases}
\end{equation}
and $\gamma=\sqrt{3}/2-1\approx0.366$ is the optimal width parameter of the square window.\cite{CottonJCP2013,Runeson2020}

The triangle window\cite{CottonJCP2016,CottonJCP2019_2} is expressed as
\begin{equation}\label{EQ:TriangleWindow}
    W_j({\bf e}) = w_{1}(e_j) \prod_{j'\ne j}^\mathcal{F} w_{0}(e_j,e_{j'}),
\end{equation}
where
\begin{equation}
    w_{1}(e) = 
    \begin{cases} 
      (2 - e)^{2-\mathcal{F}}, & 1 < e < 2 \\
      0, & \mathrm{else}
   \end{cases}
\end{equation}
and 
\begin{equation}
    w_{0}(e,e') = 
    \begin{cases} 
      1, & e' < 2 - e \\
      0, & \mathrm{else},
   \end{cases}
\end{equation}
and trajectories are assigned to state $j$ at time $t$ if $e_{j}\ge 1$ and $e_{j'}<1$ for all $j'\neq j$. The ZPE parameter in the triangle scheme is $\gamma=1/3$. The triangle window scheme for a 2-state system is depicted in Fig.~\ref{FIG1}b.

The time-dependent population at time $t$ is then calculated by applying the window function estimator to action variables $\{e_j (t)\}$
for an ensemble of trajectories. Starting from the initial diabatic state $|i \rangle$, the time-dependent population of the states $|j \rangle$ is computed with Eqn.~\ref{eqn:wignersqc}. However, by using the window function estimator, the total population is no longer properly normalized due to the fraction of trajectories that are outside of any window region at any given time.\cite{CottonJCP2013} Thus, the total population must be normalized\cite{CottonJCP2013} with the following procedure:
\begin{equation}\label{eq:popNormalize}
{{\rho}_{jj}(t)}/{\sum_{i=1}^N {\rho}_{ii}(t)}\rightarrow {\rho}_{jj}(t).
\end{equation}
This SCQ approach provides a dramatic improvement to Ehrenfest dynamics, even though they utilize the same equation of motion for the coupled electronic-nuclear DOFs.\cite{MillerFD2016,BellonziJCP2016} The SQC method allows for the elimination of known issues present in Ehrenfest dynamics, including preserving detailed balance,\cite{ParandekarJCTC2006,MillerJCP2015,BellonziJCP2016}, and this method has shown to be quite accurate in an assortment of model systems \cite{CottonJCP2014,CottonJCTC2016,CottonJCP2016,TaoJPCC2014} while only needing a few thousand trajectories for convergence.\cite{MillerFD2016,CottonJCP2013,CottonJCTC2016} As such, the SQC method is well-suited for use in on-the-fly non-adiabatic simulations of real molecular systems.

\subsection{The $\gamma$-Correction Approach} \label{SEC:ZPE-Correction}
It was recently proposed that the mapping zero-point energy should be chosen in such a way as to constrain the initial force to be composed purely from the initially occupied state,\cite{CottonJCP2019_2} which was not previously enforced using a fixed $\gamma$ in Eq.~\ref{eq:force}. This new scheme has shown to provide a significant improvement for photo-dissociation problems with coupled Morse potentials\cite{CottonJCP2019_2} and has been combined with the kinematic momentum approach\cite{CottonJCP2017} to carry out on-the-fly simulations of the methaniminium cation.\cite{HuJCTC2021}

The basic logic of this scheme is to choose an appropriate $\gamma_j$ for each state $|j\rangle$ in a given individual trajectory, such that the initial population is forced to respect the initial electronic excitation focused onto a single excited state. If the initial electronic state is $|i\rangle$, then 
\begin{equation}
     \gamma_j = e_j - \delta_{ji},
\end{equation}
or equivalently, 
\begin{equation}\label{EQ:GAMMA}
    \delta_{ji} = e_j - \gamma_j,
\end{equation}
where the $\{e_j\}$ are uniformly sampled, and then the $\gamma_j$ are chosen to satisfy Eq. \ref{EQ:GAMMA}.

These $\gamma_j$ will be explicitly used in the EOMs in Eqs.~\ref{eq:mapeqn}-\ref{eq:force}, and in particular, the nuclear forces are now
\begin{equation}
    {\bf F}=-{1\over 2}\sum_{ij}\nabla V_{ij}({\bf R})\big(p_{i}p_{j}+q_{i}q_{j}-2\gamma_{j}\delta_{ji}\big),
\end{equation}
ensuring the initial forces (at $t=0$) are simply ${\bf F}=-\nabla V_{ii}(R)$. Previously, without any adjustments to $\gamma_k$, the chosen values for $\gamma_k$ were only dependent on the windowing function itself, i.e., $\gamma_k = 0.366$ for the square Windows and $\gamma_k = 1/3$ for the triangle windows. With the above $\gamma$-correction method,\cite{CottonJCP2019_2} each individual trajectory will has its own state-specific $\gamma_j$ for state $|j\rangle$ that is completely independent of the choice of window function. The choices of the $\gamma$ parameter for different SQC approaches are summarized in Table~\ref{GAMMA}.
\begin{table}
    \centering
    \caption{Choice of $\gamma$ for state $|j\rangle$ with initial excitation on state $|i\rangle$ of each SQC method employed in this work.}
    \begin{tabular}{c c c}
    \hline
    \hline
    Method & $\gamma$-correction & $\gamma_j$\\
    \hline
    $\square$-SQC  & False & 0.366 \\
    $\Delta$-SQC & False & 1/3  \\
    $\square$-$\gamma$-SQC  & True & $e_j - \delta_{ji}$ \\
    $\Delta$-$\gamma$-SQC & True &  $e_j - \delta_{ji}$\\
    \hline
    \hline
    \end{tabular}
    \label{GAMMA}
\end{table}

Note that reformulating
$\gamma$-SQC in the adiabatic representation (such as the kinematically transformed SQC\cite{CottonJCP2017}) has been done recently to perform on-the-fly simulations.\cite{HuJCTC2021} However, formulating a quantum dynamics approach in the adiabatic representation introduces additional numerical issues as the molecular Hamiltonian in the adiabatic representation involves first and second derivative coupling elements that are typically sharply peaked around avoided crossings and become singular at a conical intersection (CI). In our previous work,\cite{SandovalJCP2018} we showed that the kinematically transformed SQC (KT-SQC), formulated in the adiabatic representation, may require substantially small time step to converge dynamics due to the presence of sharply peaked first derivative coupling elements, even though the second derivative coupling elements are removed through mathematical transformation in this formalism.\cite{CottonJCP2017} To this end, we use the Quasi-Diabatic Propagation Scheme to directly interface the diabatic $\gamma$-SQC approach with adiabatic electronic structure information. 

\subsection{Quasi-Diabatic Propagation Scheme}
In this work, we combine the $\gamma-$SQC approach formulated in the { \it diabatic} representation with {\it adiabatic} outputs of an electronic structure approach using the QD propagation scheme.\cite{MandalJCTC2018,SandovalJCP2018,ZhouJCPL2019}  

Despite recent theoretical progress,\cite{VoorhisARPC2010,SubotnikACR2015,KubasJCP2014,ZengJCTC2012,SirjoosinghJCTC2011} strict diabatic states $\{|i\rangle,|j\rangle\}$ are neither uniquely defined nor routinely available for `real' molecular systems. In contrast, it is convenient to obtain {\it adiabatic} states by solving the following eigenequation
\begin{equation}\label{eq:eigenvalue}
    \hat{V}(\hat{\bf r}, {\bf R}) |\Phi_\alpha({\bf R})\rangle = E_{\alpha}(\mathbf{R})|\Phi_\alpha({\bf R})\rangle,
\end{equation} 
where $\hat{V}(\hat{\bf r}; {\bf R})$ is the electronic part of the molecular Hamiltonian at a nuclear configuration ${\bf R}$, and $|\Phi_{\alpha}({\bf R})\rangle$ is the {\it adiabatic} state which is the eigenstate of $\hat{V}(\hat{\bf r}; {\bf R})$, with the corresponding eigenvalue $E_{\alpha}(\mathbf{R})$ referred to as the adiabatic potential energy.  


Consider a short-time propagation of the nuclear DOFs during $t\in[t_0, t_1]$, where the nuclear positions evolve from ${\bf R}(t_0)$ to ${\bf R}(t_1)$, and the corresponding adiabatic states are $\{|\Phi_{\alpha}({\bf R}(t_0))\rangle\}$ and $\{|\Phi_{\lambda}({\bf R}(t_1))\rangle\}$. The QD scheme uses the nuclear geometry at time $t_0$ as a reference geometry, ${{\bf R}_{0}}\equiv {\bf R}(t_0)$, and uses the adiabatic basis $\{|\Phi_{\alpha}({\bf R}(t_0))\rangle\}$ as the {\it quasi-diabatic} basis during this short-time propagation, such that
\begin{equation}\label{eqn:qdidea}
    |\Phi_{\alpha}({{\bf R}_{0}})\rangle\equiv|\Phi_{\alpha}({\bf R}(t_0))\rangle,~~\mathrm{for}~t\in[t_0,t_1].
\end{equation}
With the above QD basis defined independently of $R(t)$ within each propagation segment (or nuclear time step),  $\hat{V}(\hat{\bf r},{\bf R})$ in the QD basis becomes off-diagonal, while the derivative couplings vanish. Within this basis, all of the necessary diabatic quantities can be evaluated and used to propagate quantum dynamics during $t\in[t_0,t_1]$. 

The electronic Hamiltonian operator $\hat V (\hat{\bf r}, {\bf R}(t))$ in the QD basis is evaluated as
\begin{equation}\label{eqn:vijt} 
    V_{\alpha\beta}({\bf R}(t))  = \langle \Phi_\alpha ({{\bf R}_0})| \hat V ({\bf R}(t))| \Phi_\beta({{\bf R}_0})\rangle.
\end{equation}
For on-the-fly simulations, this quantity is obtained from a linear interpolation\cite{WebsterCPC1991} between $V_{\alpha\beta}({\bf R}(t_{0}))$ and $V_{\alpha\beta}({\bf R}(t_{1}))$ as follows
\begin{align}\label{eqn:interpolation}
    V_{\alpha\beta}({\bf R}(t)) & = V_{\alpha\beta}({\bf R}_{0})+\frac {(t - t_{0})}{(t_{1} - t_{0})}\big[V_{\alpha\beta}({\bf R}(t_{1})) - V_{\alpha\beta}({\bf R}_{0})\big],
\end{align} 
where $V_{\alpha\beta}({\bf R}_{0})=\langle \Phi_\alpha ({{\bf R}_0})| \hat V ({\bf R}(t_0))| \Phi_\beta ({\bf R_0})\rangle = E_{\alpha}({\bf R}(t_0))\delta_{\alpha\beta}$. The matrix elements $V_{\alpha\beta}({\bf R}(t_{1}))$ are computed as follows
\begin{equation}\label{eqn:elect2}
    V_{\alpha\beta}({\bf R}(t_{1})) =\sum_{\lambda\nu}S_{\alpha\lambda} {V}_{\lambda\nu} ({\bf R}(t_1)) S^{\dagger}_{\beta\nu},
\end{equation}
\noindent where ${V}_{\lambda\nu} ({\bf R}(t_1))=\langle \Phi_{\lambda}({\bf R}(t_{1}))| \hat {V} ({\bf R}(t_1))|\Phi_{\nu}({\bf R}(t_{1})) \rangle=E_{\lambda}({\bf R}(t_1))\delta_{\lambda\nu}$, and the overlap matrix between two adiabatic electronic states (at two different nuclear geometries) are $S_{\alpha\lambda}= \langle \Phi_{\alpha}({{\bf R}_0})|\Phi_{\lambda}({\bf R}(t_{1}))\rangle$ and $S^{\dagger}_{\beta\nu} = \langle \Phi_{\nu}({{\bf R}}(t_{1}))|\Phi_{\beta}({\bf R}_0)\rangle$. These overlap matrices are computed based on the approach outlined in Ref.~\citenum{PlasserJCTC2016}.

The nuclear gradients $\nabla V_{\alpha\beta}({\bf R}(t_{1}))\equiv \partial V_{\alpha\beta}({\bf R}(t_{1}))/\partial {\bf R}$ are evaluated as
\begin{align}\label{eqn:nucgrad}
    &\nabla V_{\alpha\beta}({\bf R}(t_{1}))=\nabla \langle \Phi_\alpha({\bf R}_{0})| \hat V ({\bf R}(t_1))| \Phi_\beta ({\bf R}_{0})\rangle \nonumber \\
    &=\langle \Phi_\alpha ({\bf R}_{0})| \nabla \hat V ({\bf R}(t_1))| \Phi_\beta ({\bf R}_{0})\rangle \\
    &=\sum_{\lambda\nu} S_{\alpha\lambda}\langle \Phi_{\lambda}({\bf R}(t_{1}))|\nabla \hat V ({\bf R}(t_1))|\Phi_{\nu}({\bf R}(t_{1})) \rangle S^{\dagger}_{\beta\nu}. \nonumber
\end{align}  
To derive the above expression, we have used the fact that $\{|\Phi_\alpha({\bf R}_0)\rangle\}$ is a {\it diabatic} basis during the $[t_0, t_1]$ propagation, allowing us to move the gradient operator to bypass $\langle \Phi_\alpha({\bf R}_{0})|$. We have also inserted the resolution of identity $\sum_{\lambda} |\Phi_{\lambda}({\bf R}(t_1))\rangle\langle \Phi_{\lambda}({\bf R}(t_1))|=1$ in the second line of the above equation and assume that the QD basis at nuclear position ${\bf R}(t_1)$ is complete. It should be noted that Eq.~\ref{eqn:nucgrad} includes derivatives with respect to all possible sources of the nuclear dependence which include those from the adiabatic potentials and the adiabatic states\cite{MandalJCP2018,MandalJPCA2019}.

During the next short-time propagation segment $t\in[t_1,t_2]$, the QD scheme adopts a new reference geometry ${{\bf R}'_{0}}\equiv {\bf R}(t_1)$ and new {\it diabatic} basis $|\Phi_{\mu}({{\bf R}'_{0}})\rangle\equiv|\Phi_{\mu}({\bf R}(t_1))\rangle$. Between $[t_0,t_1]$ propagation and $[t_1,t_2]$ propagation segments, all of these quantities will be transformed from $\{|\Phi_{\alpha}({\bf R}_{0})\rangle\}$ to $\{|\Phi_{\mu}({\bf R}'_{0})\rangle\}$ basis, using the relation 
\begin{equation}\label{eqn:basis}
    |\Phi_{\mu}({\bf R}(t_{1}))\rangle=\sum_{\alpha} \langle \Phi_{\alpha}({\bf R}(t_{0}))| \Phi_{\mu}({\bf R}(t_{1}))\rangle|\Phi_{\alpha}({\bf R}(t_{0}))\rangle.
\end{equation}
\noindent Since the mapping relation between the physical state and the singly excited oscillator state is $|\Phi_{\mu}({\bf R}(t_{1}))\rangle=a_{\mu}^{\dagger}|0\rangle=\frac{1}{\sqrt{2}}(\hat{q}_\mu + i\hat{p}_\mu)|0\rangle$ and $|\Phi_{\alpha}({\bf R}(t_{0}))\rangle=a_{\alpha}^{\dagger}|0\rangle=\frac{1}{\sqrt{2}}(\hat{q}_\alpha + i\hat{p}_\alpha)|0\rangle$, the relations for the mapping variables associated with two bases are 
\begin{align}\label{map-trans}
    \frac{1}{\sqrt{2}}(\hat{q}_\mu + i\hat{p}_\mu)|0\rangle=\sum_{\alpha}&\langle \Phi_{\alpha}({\bf R}(t_{0}))| \Phi_{\mu}({\bf R}(t_{1}))\rangle\\
    &\times\frac{1}{\sqrt{2}} (\hat{q}_\alpha + i\hat{p}_\alpha)|0\rangle.\nonumber
\end{align}
\noindent For molecular systems, one can always find a suitable choice for the basis set in order to make $\langle \Phi_{\alpha}({\bf R}(t_{0}))| \Phi_{\mu}({\bf R}(t_{1}))\rangle$ real, which guarantees that the mapping variables are transformed with the same relations as the bases. Based on this relation in Eq.~\ref{map-trans}, we transform the time-dependent mapping variables between the two consecutive QD bases as follows
\begin{subequations}\label{map-trans2}
\begin{align}
    &\sum_{\alpha} q_{\alpha} \langle \Phi_{\alpha}({\bf R}(t_{0}))|\Phi_{\mu}({\bf R}(t_{1}))\rangle \rightarrow q_{\mu}\\
    &\sum_{\alpha} p_{\alpha} \langle \Phi_{\alpha}({\bf R}(t_{0}))| \Phi_{\mu}({\bf R}(t_{1}))\rangle \rightarrow p_{\mu}
\end{align}
\end{subequations}
\noindent When performing the transformation in Eq.~\ref{eqn:basis} and Eq.~\ref{map-trans2}, the eigenvectors maintain their mutual orthogonality subject to a very small error when they are expressed in terms of the previous basis due to the incompleteness of the basis.\cite{GranucciJCP2001,PlasserJCP2012} Nevertheless, the orthogonality remains to be well satisfied among  $\{|\Phi_{\alpha}({\bf R}(t_0))\rangle\}$ or $\{|\Phi_{\lambda}({\bf R}(t_1))\rangle\}$. This small numerical error generated from each step can, however, accumulate over many steps and cause a significant error at longer times, leading to non-unitary dynamics.\cite{GranucciJCP2001,PlasserJCP2012} This potential issue can be easily resolved by using orthonormalization procedure among the vectors of the overlap matrix $\langle \Phi_{\alpha}({\bf R}(t_{0}))| \Phi_{\mu}({\bf R}(t_{1}))\rangle$, as been done in our previous work\cite{MandalJCP2018} for simulating photoinduced charge transfer dynamics. Here, we perform the L\"owdin orthogonalization procedure\cite{LowdinJCP1950} as commonly used in the the local diabatization approach\cite{GranucciJCP2001} to ensure this. 

As the nuclear geometry closely follows the reference geometry throughout the  propagation, the QD basis forms a convenient and compact basis. Note that, in principle, one needs infinite crude adiabatic states $\{|\Phi_\alpha({\bf R}_0)\rangle\}$ to represent the time-dependent electronic wavefunction, because the electronic wavefunction could change rapidly with the motion of the nuclei, and the crude adiabatic basis is only convenient when the reference geometry ${\bf R}_{0}$ is close to the nuclear geometry ${\bf R}$. By dynamically updating the basis in the QD scheme, the time-dependent electronic wavefunction is expanded with the ``moving crude adiabatic basis"\cite{IzmaylovJCP2018} that explores the most relevant and important parts of the Hilbert space, thus requiring few states for quantum dynamics propagation. 

Thus, the QD representation provides several unique advantages over the strict diabatic or adiabatic representation for quantum dynamics propagation. On one hand, the QD basis is constructed from the crude adiabatic basis, which can be easily obtained from any commonly used electronic structure calculation. On the other hand, the diabatic nature of the QD basis makes derivative couplings explicitly vanish and allows using any diabatic dynamics approaches to perform on-the-fly propagation. Further, the QD scheme ensures a stable propagation of the quantum dynamics compared to directly solving it in the adiabatic representation. This is due to the fact that directly solving electronic dynamics in the adiabatic state requires the non-adiabatic coupling $\langle \Phi_\beta({\bf R}(t))|{\partial\over{\partial t}}\Phi_\alpha({\bf R}(t))\rangle={\bf d}_{\beta\alpha}({\bf R})\cdot\dot{\bf R}$, which might exhibit highly peaked values and cause large numerical errors\cite{MeekJPCL2014, JainJCTC2016} when using a linear interpolation scheme.\cite{SharonJCP1994} The QD scheme explicitly alleviates this difficulty by using the well behaved transformation matrix elements $\langle \Phi_\beta({\bf R}(t_1))|\Phi_\alpha({\bf R}(t_2))\rangle$ instead of $\langle \Phi_\beta({\bf R}(t))|{\partial\over{\partial t}}\Phi_\alpha({\bf R}(t))\rangle$. 

Note that the SQC approach has been derived in the adiabatic representation,\cite{CottonJCP2017} which contains the derivative of the derivative coupling $\nabla{\bf d}_{\alpha\beta}({\bf R})$ in the equation of motion. With kinematic momentum transform, it explicitly eliminates the presence of $\nabla{\bf d}_{\alpha\beta}({\bf R})$ term in the nuclear force (instead of ignoring it), and the nuclear EOM of the KM-SQC approach is equivalent to the nuclear forces in QD-SQC.\cite{SandovalJCP2018} On the other hand, the KM-SQC EOMs still explicitly contain $\mathbf{d}_{\beta\alpha}(\mathbf{R})$ (through the presence of $\langle \Phi_\beta({\bf R}(t))|{\partial\over{\partial t}}\Phi_\alpha({\bf R}(t))\rangle$, which could lead to numerical instabilities when these derivative couplings are highly peaked. This has been extensively investigated in Ref.~\citenum{SandovalJCP2018} using Tully's avoid crossing model, which has a very narrow derivative coupling, such that it can drastically change on a time-scale that is shorter than the nuclear time step $dt$. When using a large $dt$ in the KM-SQC approach, the nuclear position can encounter drastically different values of the derivative coupling from one step to another that allow for a discontinuous spike at a CI or even completely step over it,\cite{MeekJCPL2014} resulting in different long-time populations and oscillatory behavior of errors. Thus, the approaches that explicitly require derivative couplings (and those which use a simple linear interpolation scheme for obtaining them, as we have previously implemented for the KM scheme) either encounter numerical challenges or start to accumulate numerical errors.\cite{MeekJCPL2014} The QD scheme, on the other hand, provides more accurate results even when using a relatively larger time step $dt$, simply because the QD schemes only requires the well-behaved transformation matrix elements $\langle \Phi_1({\bf R}(t_0))|\Phi_2({\bf R}(t_1))\rangle$ instead of the highly peaked derivative coupling ${\bf d}_{12}({\bf R})$. That being said, there may be good alternative approaches to achieve the same attractive features for dynamics propagation, such as those recently developed norm-preserving interpolation schemes.\cite{MeekJCPL2014, JainJCTC2016} The QD scheme is perhaps, still the most straightforward one that allows robust dynamical propagation and enables a seamless interface between the diabatic quantum dynamics approach (such as SQC) and adiabatic electronic structure calculations.

\subsection{Computational Details}
Non-adiabatic molecular dynamics simulations based on the QD-$\gamma$-SQC approach are performed using an in-house modified version\cite{ZhouJCPL2019} of the SHARC non-adiabatic molecular dynamics code, interfaced to the MOLPRO electronic structure package.\cite{mai_sharc21_2019,werner_molpro_2012} 
On-the-fly electronic structure calculations are performed at the level of complete active space self-consistent field (CASSCF) approach, with 3SA-CASSCF(2,2)/6-31G* and 2SA-CASSCF(6,6)/6-31G* level of theory for ethylene and fulvene, respectively.\cite{IbelePCCP2020} The CAS self-consistent calculation is performed over three lowest adiabatic states for ethylene and over two lowest adiabatic states for fulvene, whereas the quantum dynamics for both molecules are only confined in the $\{\mathrm{S}_0, \mathrm{S}_1\}$ subspace. All of the energies and gradients are computed at this level of electronic structure theory. The default accuracy for both the nuclear gradients and non-adiabatic vectors is $10^{-7}$ a.u.; when this criterion is not satisfied, a maximum of 900 additional wavefunction optimization iterations are used to make sure the convergence of $10^{-4}$ a.u. is reached. All of the electronic structure calculations performed during our quantum dynamics simulations converge successfully under the above criteria.

\begin{figure}
 \centering
    \begin{minipage}[h]{1.0\linewidth}
     \centering
     \includegraphics[width=\linewidth]{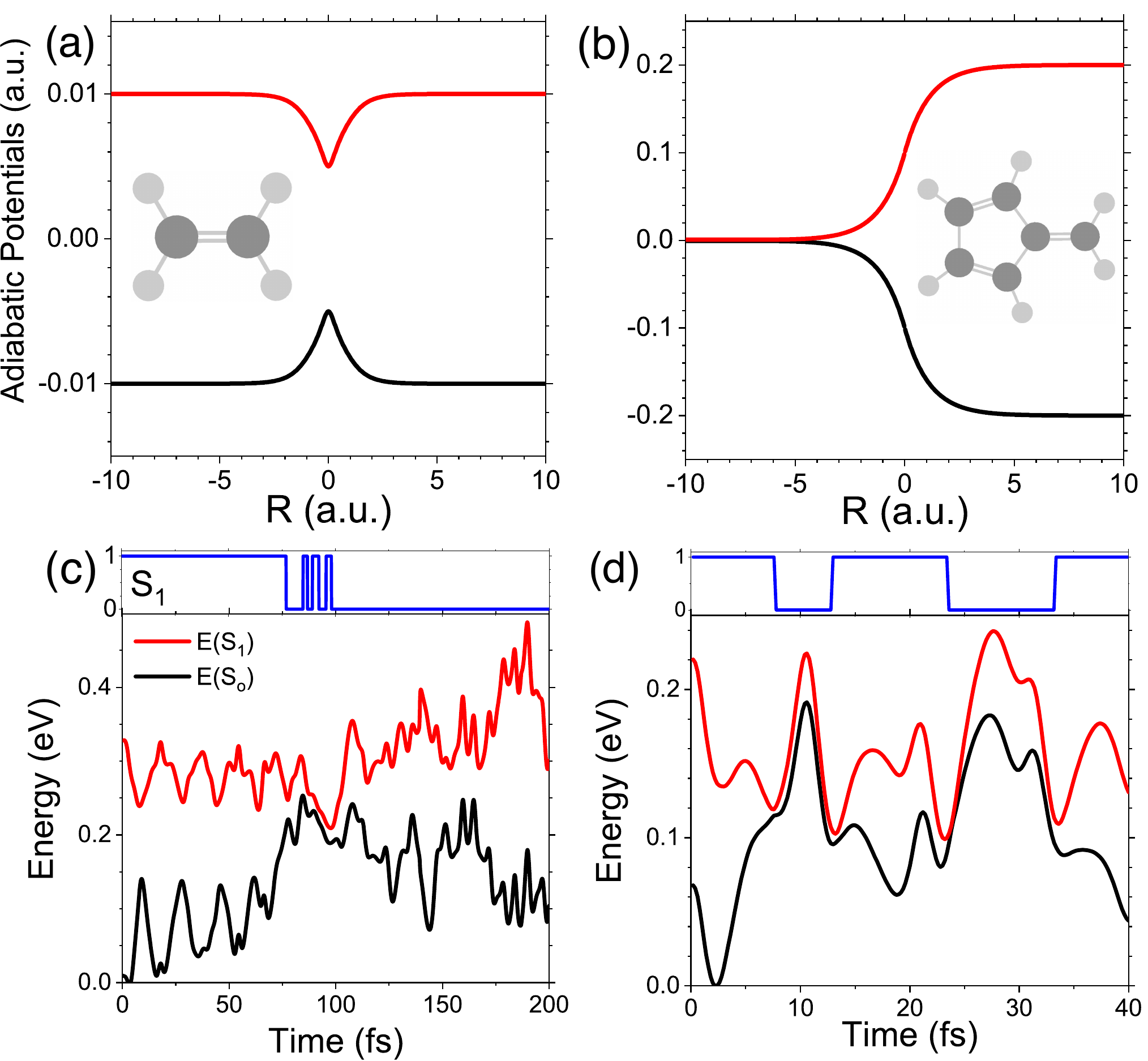}
    \end{minipage}%
   \caption{Adiabatic potentials for (a) Tully's model I (a single avoided crossing) and (b) Tully's model III (an extended region of coupling with a reflective barrier). The molecular structures of the {\it ab-initio} Tully models are depicted in the insets. Along a single QD-$\gamma$-SQC trajectory, the population of the $\mathrm{S}_1$ (blue) state and the energies of the $\mathrm{S}_0$ (black) and $\mathrm{S}_1$ (red) states as functions of time for (c) ethylene and (d) fulvene are presented.}
\label{FIG2-Tully}
\end{figure}

The initial Wigner distribution is sampled from the ground vibrational state $\nu=0$ on the ground electronic state $\mathrm{S}_0$, where the normal mode frequencies (in the harmonic approximation) are calculated based on the approach outlined in Ref.~\citenum{DahlJCP1988,SchinkeBook}, as implemented in the SHARC package. The normal mode frequencies are computed at the level of MP2/6-31++G** with the MOLPRO package, with the optimized structure obtained at the same level of electronic structure theory for the ground state $|S_{0}({\bf R})\rangle$. In particular, the nuclear density $\rho_W(\tilde{\bf{R}},\tilde{\bf{P}})$ in terms of the molecular normal-mode frequencies $\{\tilde{\omega}_{k}\}$ and phase space variables $\{\tilde{\bf{R}},\tilde{\bf{P}}\}$ is given as\cite{Tannor} 
\begin{equation}\label{wig}
\rho_\mathrm{W}(\tilde{\bf{R}},\tilde{\bf{P}})\propto \prod_{k=1}^{\mathcal{N}}\exp[{-\tanh(\frac{\beta\hbar\tilde{\omega}_k}{2})(\frac{m\tilde{\omega}_k}{\hbar}\tilde{{R}}_{k}^2+\frac{1}{m_{k}\tilde{\omega}_{k} \hbar}\tilde{P}_{k}^2)}].
\end{equation}
The initial distribution $\{{\bf{R}},{\bf{P}}\}$ is then obtained by transforming $\{\tilde{\bf{R}},\tilde{\bf{P}}\}$ from the normal mode representation to the primitive coordinates using the unitary transformation that diagonalizes the Hessian matrix. 

A total number of 500 trajectories are used in the QD-$\gamma$-SQC simulations to achieve converged population dynamics. A rough convergence of the population dynamics can be already achieved within 50-100 trajectories for both molecules. The nuclear time step used in the QD-$\gamma$-SQC is $dt=0.1$ fs, with 200 electronic time steps for the mapping variables' integration during each nuclear time step. The electronic structure calculations are performed only at the nuclear time step.

The overlap matrix of CAS wavefunctions between two successive nuclear time steps is calculated by using the approach outlined in Ref.~\citenum{PlasserJCTC2016}, as implemented in the program \textsc{wfoverlap}. The
random phases generated from electronic structure calculations for the eigenfunctions are carefully calculated and accumulated. To ensure the orthonormalization among the vectors of the overlap matrix $\langle \Phi_{\alpha}({\bf R}(t_{0}))| \Phi_{\mu}({\bf R}(t_{1}))\rangle$, we perform the L\"owdin orthogonalization. \cite{GranucciJCP2001} All of the above routines were used as implemented in the SHARC program\cite{SHARC,SHARC2}.

In this work, we chose two molecular systems (i) ethylene and (ii) fulvene, that were previously investigated as the {\it ab-initio} analogies to Tully's curve crossing models,\cite{IbelePCCP2020} as illustrated in Fig.~\ref{FIG2-Tully}a,b. In panel (c), the time-dependent adiabatic potential energies of the ethylene molecule along a single nuclear trajectory for $\mathrm{S}_0$ and $\mathrm{S}_1$ states are presented. In the top sub-panel, the excited state population (blue) computed with the QD-$\gamma$-SQC approach is plotted, which shows an oscillation of the population in the the avoided crossing region that eventually relaxes down to the ground state. In panel (d), the time-dependent adiabatic potential energies of the fulvene molecule are shown and exhibit many instances of potential energy barriers within regions of near degeneracy between states, leading to reflection given a sufficiently low momentum, similar to the Tully's model III (which is described canonically by a single spatial coordinate).\cite{TullyJCP1990} The previously encountered avoided crossing is then visited again where the population transfers back into the excited state. This occurs twice (within the allotted time-frame) in the single trajectory presented in panel (d), which gives rise to the important features found in the population dynamics discussed in more detail later. These two molecules provide rich physics of two vastly different levels of dynamical complexity. It is important to note that the features which the Tully models present are with respect to a single {\it nuclear coordinate}, but in the {\it ab initio} molecular models the analogy is represented through the {\it time-dependent} adiabtic potential energies, demonstrated in Fig.~\ref{FIG2-Tully}c,d.

\section{Results and Discussion}
\begin{figure}
 \centering
  \begin{minipage}[h]{1.0\linewidth}
     \centering
     \includegraphics[width=\linewidth]{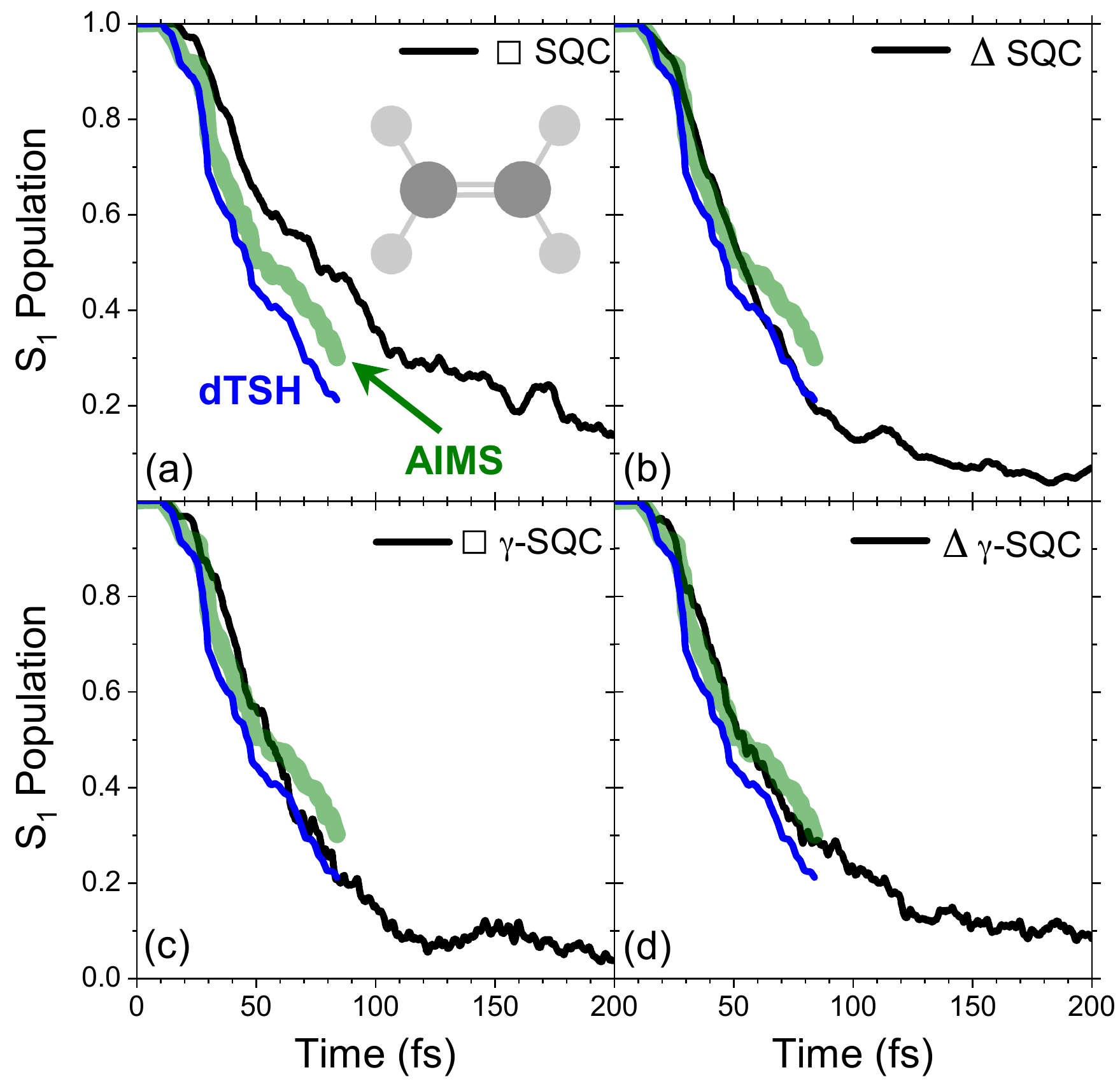}
       \end{minipage}%
   \caption{Population dynamics of the $\mathrm{S}_1$ state in ethylene,  using the square (a,b) and triangle (c,d) windowing schemes. Panels (a,c) utilize the original SQC method with fixed zero-point energy (ZPE) parameter $\gamma$, while (b,c) were computed using the trajectory-adjusted ZPE parameter $\gamma$.}
\label{FIG:Ethylene_AllMethods}
\end{figure}

Fig. \ref{FIG:Ethylene_AllMethods} presents the adiabatic population dynamics of ethylene upon photo-excitation to the $S_1$ state, obtained from various SQC approaches (black solid lines), compared to the {\it ab-initio} multiple spawning (AIMS) (thick green lines), an approximate Gaussian wavepacket-based non-adiabatic method, which is used as the benchmark result. The decoherence-corrected trajectory surface hopping (dTSH) approach (blue lines) is also presented for comparison. Both AIMS and dTSH results are directly adapted from Ref.~\citenum{IbelePCCP2020} 

Fig.~\ref{FIG:Ethylene_AllMethods}a presents the results obtained from the original\cite{CottonJPCA2013,CottonJCP2013} SQC approach using the square window scheme, where the mapping ZPE $\gamma=0.366$ is kept as a constant for all states and trajectories. Fig.~\ref{FIG:Ethylene_AllMethods}b presents the results obtained from the SQC approach using the triangle window scheme\cite{CottonJCP2016} with a fixed $\gamma=1/3$. Both approaches provide reasonably accurate non-adiabatic dynamics compared to the AIMS results, as well as to the dTSH simulations. In particular, the $\square$-SQC method seems to show an increased relaxation time compared to AIMS, whereas the triangle window scheme presented in panel (b) is more accurate for simulating the Ethylene non-adiabatic dynamics, compared to the square window scheme presented in panel (a). This trend is in an agreement with the empirical results of the recent numerical tests of both window schemes with a wide range of diabatic models,\cite{CottonJCP2016,CottonJCP2019_1} especially for models with weak non-adiabatic coupling.\cite{CottonJCP2016}

Fig.~\ref{FIG:Ethylene_AllMethods}c,d presents the ZPE-corrected QD-$\gamma$-SQC dynamics, obtained with the square window (panel c) and the triangle window (panel d). For the square window scheme, we find that the $\gamma$-SQC approach (panel c) provides much better agreement with the AIMS benchmark compared to the original SQC method (panel a). For the triangle window scheme, we find very similar short-time relaxation curves for the $\Delta$-SQC (panel b) method compared to $\Delta$-$\gamma$-SQC (panel d). For the particular case of the ethylene photo-dissociation dynamics, the ZPE correction does not further improve the results when the triangle window is used, similar to the recent work that utilized the kinematic momentum formulation of SQC.\cite{HuJCTC2021} With both $\Delta$-SQC (panel b) and  $\Delta$-$\gamma$-SQC (panel d), we see a near quantitative agreement with AIMS, even slightly out-performing the commonly used dTSH (blue).

\begin{figure}
 \centering
  \begin{minipage}[h]{1.0\linewidth}
     \centering
     \includegraphics[width=\linewidth]{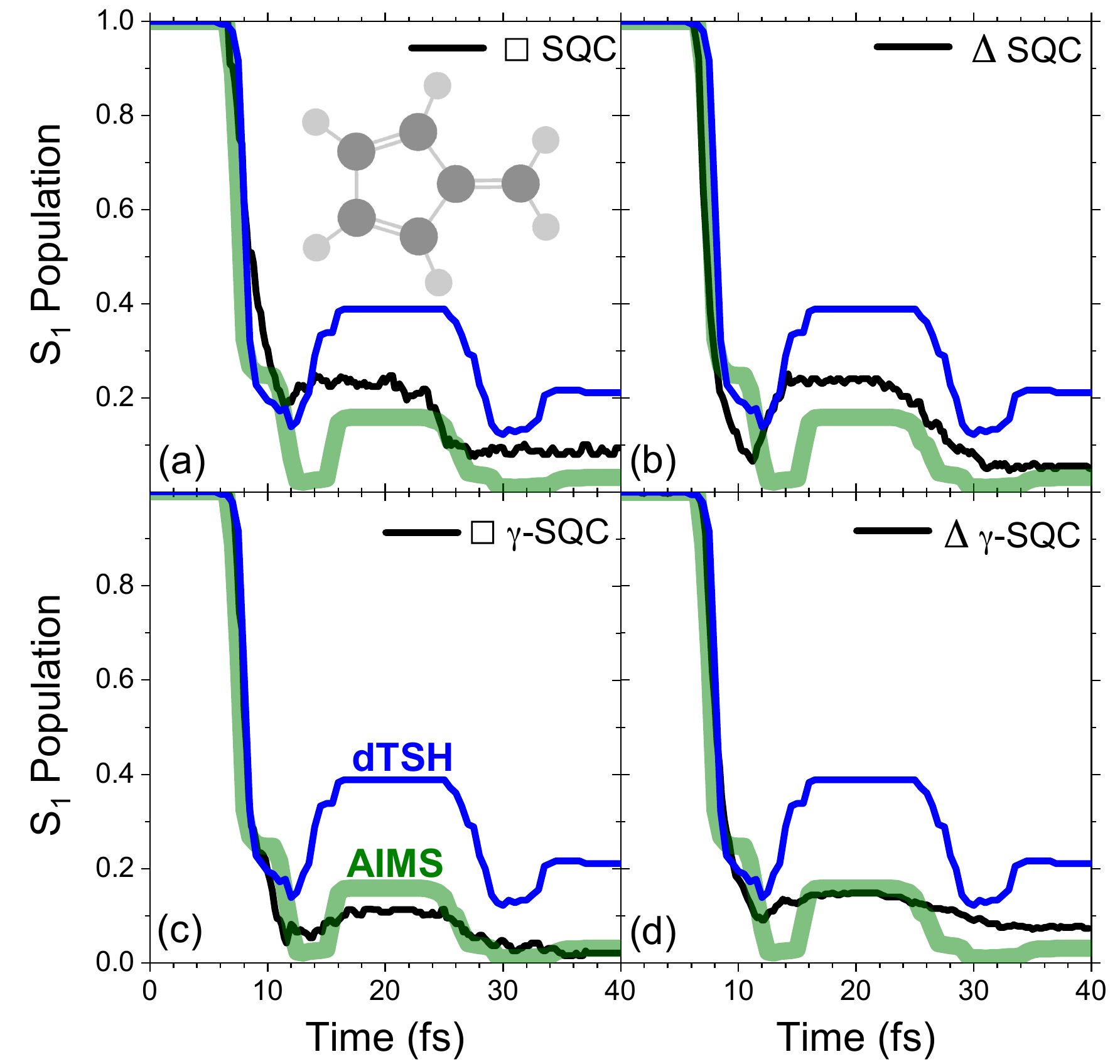}
       \end{minipage}%
   \caption{$\mathrm{S}_1$ Population dynamics for fulvene using the square (a,b) and triangle (c,d) windowing schemes. Panels (a,c) utilize the original SQC method with fixed zero-point energy (ZPE) parameter $\gamma$, while (b,c) were computed using the trajectory-adjusted ZPE parameter $\gamma$.}
\label{FIG:Fulvene_AllMethods}
\end{figure}

Fig.~\ref{FIG:Fulvene_AllMethods} presents the non-adiabatic photo-relaxation dynamics of fulvene, which has recently been proposed as a molecular example of Tully's model III.\citenum{IbelePCCP2020} In fulvene, there exists a so-called slanted CI whereby the wavepacket becomes reflected and re-interacts with this CI at nearly periodic times later ($\sim 20$-fs intervals), leading to the break down of the mixed quantum-classical methodologies due to wavepacket bifurcation as well as the added effects of encircling the CI. These non-adiabtic methods assume a single Gaussian wavepacket basis for describing the nuclear motion -- higher-order modes or multiple Gaussian wavepackets are needed to fundamentally capture these effects. For the single SQC trajectory presented in Fig.~\ref{FIG2-Tully}d, one can see that the population (blue solid line) resides in the ground state while traversing the CI but jumps back to the excited state after the interaction. Eventually, the CI will lead to the permanent relaxation to the ground state.

Fig.~\ref{FIG:Fulvene_AllMethods}a presents the results obtained from $\square$-SQC, and Fig.~\ref{FIG:Fulvene_AllMethods}b presents the results from $\Delta$-SQC. Both are providing accurate dynamics at the short time compared to AIMS, including the small shoulder in $S_1$ population at $t\sim10$ fs as well as the subsequent plateaus in ${\mathrm{S}_1}$ population. dTSH results (blue) also provides a reasonable description for the short-time dynamics. However, at a longer time for $t>10$ fs, both SQC and dTSH deviate from the AIMS results, although the SQC approach (both the square and the triangle windows) outperform dTSH. Even using different decoherence schemes in dTSH seems unable to provide further improvement, as shown in Ref.~\citenum{Vindel-Zandbergen2021}. 

Fig.~\ref{FIG:Fulvene_AllMethods}c,d presents the SQC dynamics with the $\gamma$ correction. Contrary to the ethylene case, the SQC dynamics is significantly improved upon the $\gamma$-correction for both window schemes. In particular, the $\square$-$\gamma$-SQC method (panel c) provides the most quantitative accuracy among all of the SQC-related methods, capturing both the the initial inversion of population as well as the population plateau around 20 fs and finally the tail plateau around 40 fs, slightly outperforming the $\Delta$-$\gamma$-SQC method (panel d). Both $\gamma$-SQC methods greatly outperform the dTSH simulation, as there is a severe underestimation of the population transfer in the dTSH method.\cite{IbelePCCP2020} 

\section{Conclusions}
In this work, we use the quasi-diabatic propagation scheme to directly interface the diabatic symmetric quasi-classical (SQC) approach with the electronic zero-point energy correction (the $\gamma$ correction)\cite{CottonJCP2019_2} and the CASSCF on-the-fly electronic structure calculations to propagate {\it ab-initio} non-adiabatic dynamics. We have performed simulations for two recently suggested molecular models, ethylene and fulvene, that are closely related to the well-known Tully's simple curve crossing models. We have shown that the $\gamma$-SQC method based on the trajectory-adjusted electronic zero-point energy in classical Meyer-Miller vibronic dynamics provide very accurate non-adiabatic population dynamics when comparing to \textit{ab initio} multiple spawning (AIMS), and even outperforms the widely-used adiabatic decoherence-adjusted surface hopping (dTSH) method. Specifically, for the fulvene molecule (which is an molecular analog of Tully's model III, we found that the $\gamma$-correction significantly improved the accuracy of the original SQC approach, for both the square and triangle window schemes. These calculations provide useful and non-trivial tests to systematically investigate the numerical performance of various diabatic quantum dynamics approaches, going beyond the simple diabatic model systems that have historically been the major workhorse in the quantum dynamics field. At the same time, these available benchmark studies will also likely foster the development of new quantum dynamics approaches.

\section{Acknowledgments}
This work was supported by the National Science Foundation CAREER Award under the Grant No. CHE-1845747, as well as by a Cottrell Scholar award (a program by Research Corporation for Science Advancement). B. M. W. appreciates the support from the graduate fellowship of the Department of Physics and Astronomy at the University of Rochester. A.M. appreciates the support from the ACS Chemical Computing Group Excellence Award for Graduate Students. P.H. appreciates the support from the ACS COMP OpenEye Outstanding Junior Faculty Award. Computing resources were provided by the Center for Integrated Research Computing (CIRC) at the University of Rochester.

\section*{Availability of Data}	
The data that support the findings of this study are available from the corresponding author upon a reasonable request.

%

\end{document}